\documentclass[%
reprint,
superscriptaddress,
 amsmath,amssymb,
 aps,
 prl,
 lengthcheck,%
]{revtex4}
\pdfoutput=1

\usepackage{mathptmx}

\usepackage[squaren]{SIunits}  
\usepackage{graphicx}    

\usepackage{hyperref}

\newcommand{\bra}[1]{\langle #1|}
\newcommand{\ket}[1]{|#1\nolinebreak[4]\rangle}

\begin{document}


\title{Entangling remote nuclear spins linked by a chromophore}




\author{Marcus Schaffry}
\author{Vasileia Filidou}
\affiliation{Department of Materials, University of Oxford, Parks Road, Oxford OX1 3PH, UK}
\author{Steven D. Karlen}
\affiliation{Department of Materials, University of Oxford, Parks Road, Oxford OX1 3PH, UK}
\affiliation{CRL, Department of Chemistry, University of Oxford, Oxford OX1 3TA, UK}
\author{Erik M. Gauger}
\affiliation{Department of Materials, University of Oxford, Parks Road, Oxford OX1 3PH, UK}

\author{Simon C. Benjamin}
\affiliation{Department of Materials, University of Oxford, Parks Road, Oxford OX1 3PH, UK}
\affiliation{Centre for Quantum Technologies, National University of Singapore, 3 Science Drive 2, Singapore 117543}

\author{Harry L. Anderson}
\affiliation{CRL, Department of Chemistry, University of Oxford, Oxford OX1 3TA, UK}
\author{Arzhang Ardavan}
\affiliation{CAESR, The Clarendon Laboratory, Department of Physics, University of Oxford, OX1 3PU, UK}
\author{G. Andrew D. Briggs}
\affiliation{Department of Materials, University of Oxford, Parks Road, Oxford OX1 3PH, UK}

\author{Kiminori Maeda}
\author{Kevin B. Henbest}
\affiliation{PTCL, Department of Chemistry, University of Oxford, South Parks Road, Oxford OX1 3QZ, UK}
\affiliation{CAESR, ICL, Department of Chemistry, University of Oxford, South Parks Road, Oxford OX1 3QR, UK}

\author{Feliciano Giustino}
\affiliation{Department of Materials, University of Oxford, Parks Road, Oxford OX1 3PH, UK}

\author{John J. L. Morton}
\affiliation{Department of Materials, University of Oxford, Parks Road, Oxford OX1 3PH, UK}
\affiliation{CAESR, The Clarendon Laboratory, Department of Physics, University of Oxford, OX1 3PU, UK}

\author{Brendon W. Lovett}
\affiliation{Department of Materials, University of Oxford, Parks Road, Oxford OX1 3PH, UK}

\date{\today}

\begin{abstract}
Molecular nanostructures may constitute the fabric of future quantum technologies, if their degrees of freedom can be fully harnessed. Ideally one might use nuclear spins as low-decoherence qubits and optical excitations for fast controllable interactions. Here, we present a method for entangling two nuclear spins through their mutual coupling to a transient optically-excited electron spin, and investigate its feasibility through density functional theory and experiments on a test molecule. From our calculations we identify the specific molecular properties that permit high entangling power gates under simple optical and microwave pulses; synthesis of such molecules is possible with established techniques.
\end{abstract}


\maketitle

Molecules are promising building blocks for quantum technologies, due to their reproducible nature and ability to self-assemble into complex structures. However, the need to control interactions between adjacent qubits represents a key challenge~\cite{Lehmann.Gaita-Ario.ea2007Spinqubitswith,Benjamin.Bose2003QuantumComputingwith,Ashhab.Niskanen.ea2008Interqubitcouplingmediated,Gauger.Rohde.ea2008Strategiesentanglingremote}. We here describe a method for optical control of a nuclear spin-spin interaction that presents several advantages over conventional NMR quantum processors: the spin-spin interaction can be switched, the gates are faster, and the larger energy transitions facilitate polarization transfer onto the nuclear spins. After explaining the theory behind our method, we present an experimental study of a test molecule, and show with density functional theory that an entangling gate could be achieved. 

\begin{figure}[h!]
\includegraphics{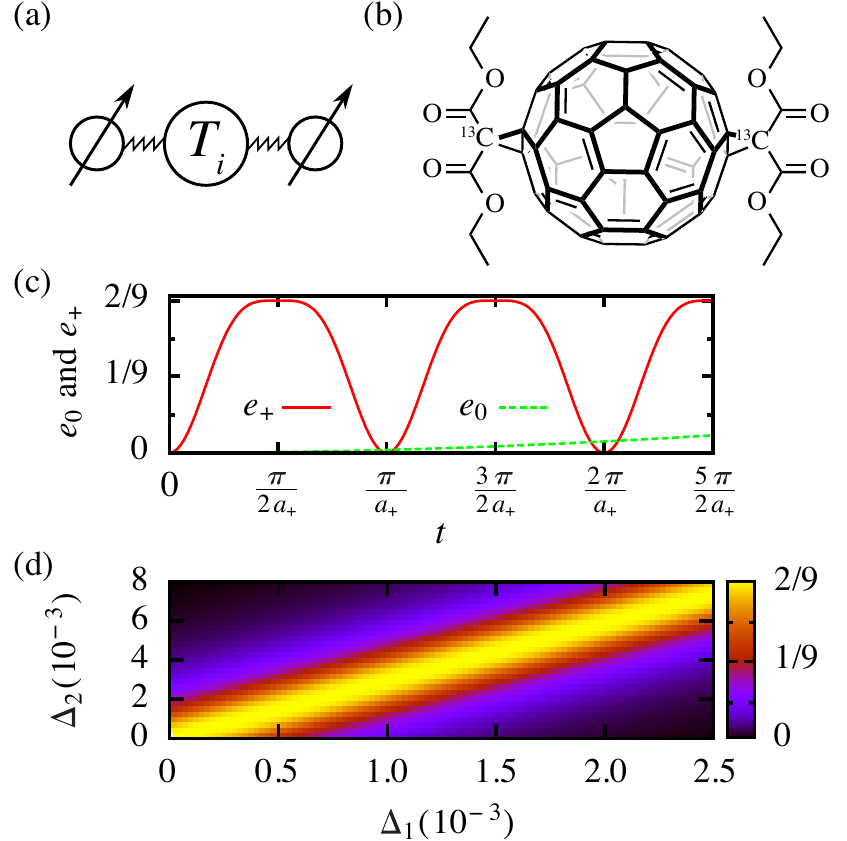} 
  \caption{(Color online) (a) A triplet state of an electron and hole couples two nuclear spin qubits. The triplet can be $|T_+\rangle=|\uparrow_e\uparrow_h\rangle$,  $|T_{0}\rangle=\frac{1}{\sqrt{2}}\bigl(|\downarrow_e \uparrow_h\rangle+|\uparrow_e\downarrow_h\rangle \bigr)$ or $|T_{-}\rangle=|\downarrow_e\downarrow_h\rangle$. (b) An example molecular embodiment of such a model system is bis-diethyl malonate fullerene. (c) Entangling power of the time evolution operators $U_+$ (solid curve) and $U_0$ (dashed curve); here $a_+\approx 32 a_0$. (d) Maximal attainable entangling power $m_{-}$ for $U_-$ as the symmetry of the two nuclear spins is reduced. $\Delta_1 = \tfrac{|\omega_{n'}-\omega_n|}{\omega_n}$;  $\Delta_2 = \tfrac{|A'-A|}{A}$;  $A=\unit{2.5}{\mega\hertz}$; $D=\unit{-296}{\mega\hertz}$; $\omega_e =\unit{9.6}{\giga\hertz}$; $\omega_n =\unit{3.7}{\mega\hertz}$.}  \label{fig:entangling-power}
\end{figure}

We consider two nuclear spin qubits labelled $n$ and $n'$ and one mediator $e$. The mediator possesses a paramagnetic excited state $\ket{e}$ with spin one character and a diamagnetic, spinless ground state $\ket{0}$. The two nuclear qubits do not interact with each other directly, but both couple to the excitation via an isotropic hyperfine (HF) coupling with generally unequal strengths $A$ and $A'$, see Fig.~\ref{fig:entangling-power} (a).  The Hamiltonian in a magnetic field is given by ($\hbar = 1$):
\begin{multline}
      \label{eq:1}
H=-\omega_n S_{z,n}- \omega_{n'} S_{z,n'}+\ket{e} \bigl( \omega_e S_{z,e} +\omega_0 \bigr) \bra{e} \\
+\ket{e} \bigl( A  \mathbf{S}_n\cdot \mathbf{S}_e + A' \mathbf{S}_{n'}\cdot \mathbf{S}_e   + D S_{z,e}^2 \bigr) \bra{e}\quad,  
\end{multline}
where $S_{z,i}$ and $\mathbf{S}_{i}$ are the Pauli spin operators and $\omega_i$ denotes the respective Zeeman splittings ($i = n, n', e$). $D$ is the zero-field-splitting (ZFS). $\omega_0$ denotes the optical frequency corresponding to the creation energy of the excitation. 

Let us first analyze the case of a symmetric (homonuclear) system with $\omega_n=\omega_{n'}$ and $A=A'$. Using degenerate perturbation theory and assuming that the electronic Zeeman splitting is much larger than that of the nuclei, $\omega_n \ll \omega_e$ and the ZFS, $D\ll \omega_e$,  we obtain an effective Hamiltonian by approximating $H_{\text{sym}}:=\bra{e}H \ket{e}$ in the following way:
\begin{equation}
  \label{eq:3}
 H_{\text{sym}} \approx V^\dagger \text{diag}(E_1,\ldots,E_{12}) V =: H_{\text{sym,eff}}  \quad,
\end{equation}
where $V =  ( \ket{\psi_1} \ket{\psi_2} \cdots \ket{\psi_{12}} )$ is the matrix of the approximate eigenvectors up to first order and the $E_i$ ($i=1,\ldots,12$) are the eigenenergies up to second order. This reveals that the time evolution of the entire system can be decomposed into three different unitary actions on the nuclear spin, one for each spin projection of the excitation $\ket{T_i}$ ($i=-,0,+$):
\begin{equation}
U(t) \approx  \exp\left(-i H_{\text{sym,eff}} t\right)  =U_{-}(t)\oplus U_0(t)\oplus U_{+}(t)  \quad. \label{eq:unitary}
\end{equation}
Moreover, only the states $\ket{T_i\downarrow\uparrow}$ and $\ket{T_i \uparrow\downarrow}$ are coupled, with a matrix element $a_i$:
\begin{equation}
  \label{eq:4}
  a_{\pm} = \frac{2 A^2}{\mp D + 2 \omega_e + 4 \omega_n};\, a_0 = a_+ - a_- = O\left(\frac{DA^2}{\omega_e^2}\right). 
\end{equation}

This coupling can take an initial product state to an entangled state of nuclear spins and we can quantify the degree of entanglement by employing the entangling power $e_i$ of a unitary $U_i$. $e_i$ is the mean linear entropy produced by the unitary acting on a uniform distribution of all pure product states. A maximally entangling gate has $e_i=\frac{2}{9}$ \cite{Zanardi.Zalka.ea2000Entanglingpowerof}. For Eq.~\eqref{eq:unitary} we get
\begin{equation}
  \label{eq:ent-power}
  e_{i}(t)=\frac{1}{9} \bigl(3 + \cos(2a_it) \bigr) \sin^2(a_it) \text{ for } i=-,0,+\quad.
\end{equation}
Hence $U(t)$ is maximally entangling at odd integer multiples of $t=\tfrac{\pi}{2a_i}$. $e_0(t)\approx 0$ for $t<\tfrac{\pi}{2a_{\pm}}$ since $a_0 \ll a_{\pm}$ (see Fig.~\ref{fig:entangling-power}). 

A \emph{protocol} for creating and preserving entanglement of the nuclear qubits is thus to start in the singlet ground state of the electron and to initialize the nuclear spins, e.g. in the state $\ket{\downarrow\uparrow}$. Polarizing nuclear spins in an NMR experiment is very challenging, however in this case one could initialize the nuclear spin by transferring  the electronic spin polarization. A laser pulse then creates the electron triplet. Our experiments (see below) revealed that the state of the excitation depends on the orientation of the molecule to the static magnetic field. As a start we assume it is in $\ket{T_0}$. With a strong microwave pulse we flip the state of the excitation, say to the $\ket{T_+}$ state, and then wait a time $t=\tfrac{\pi}{2a_+}$. A second microwave pulse is then used to flip the excitation back into $\ket{T_0}$. If the lifetime of the excitation $\tau$ is bounded by
\begin{equation}
  \label{eq:2}
\tfrac{\pi}{2a_{\pm}} < \tau \ll \tfrac{\pi}{2a_0}  
\end{equation}
the nuclear spins remain entangled for the duration of nuclear spin coherence \footnote{An optical de-excitation pulse removes the need for the upper bound, though we exploit this same bound for our model of spontaneous emission.}, because the electronic ground state is spin silent.

Such a scheme could be demonstrated using an ensemble of molecules, so long as we address two challenges. First, spontaneous optical decay must be taken into account. In general, the decay process gives rise to {\em distinguishable} photons, destroying coherence between the various spin states. However, in the limit of Eq.~\eqref{eq:2}, the states $\ket{T_0\uparrow\downarrow}$ and $\ket{T_0\downarrow\uparrow}$ decay producing indistinguishable photons, and, therefore, nuclear spin coherence in this space will survive the excitation decay. Specifically, $a_0$ is much smaller than the natural linewidth of the optical transition (unless $A$ assumes a very large value). Second the qubits must be individually addressable to verify entanglement generation with state tomography. This requires an asymmetry in $\omega_{n,i}$. Generally, the eigenspectrum of an asymmetric molecule is not degenerate, and this tarnishes our scheme since $e_i$ is then close to zero for all times. However, perturbation theory shows that under the condition $A'-A \approx \pm 2 (\omega_{n'}-\omega_n)$, we can recover a degenerate eigenspectrum for the $\ket{T_{\pm}}$ states even for an asymmetric molecule. This leads to a very similar behaviour for the entangling power as in Eq.~\eqref{eq:ent-power}, see Fig.~\ref{fig:entangling-power} (d).

\begin{figure}[h!]
  \centering
  \includegraphics{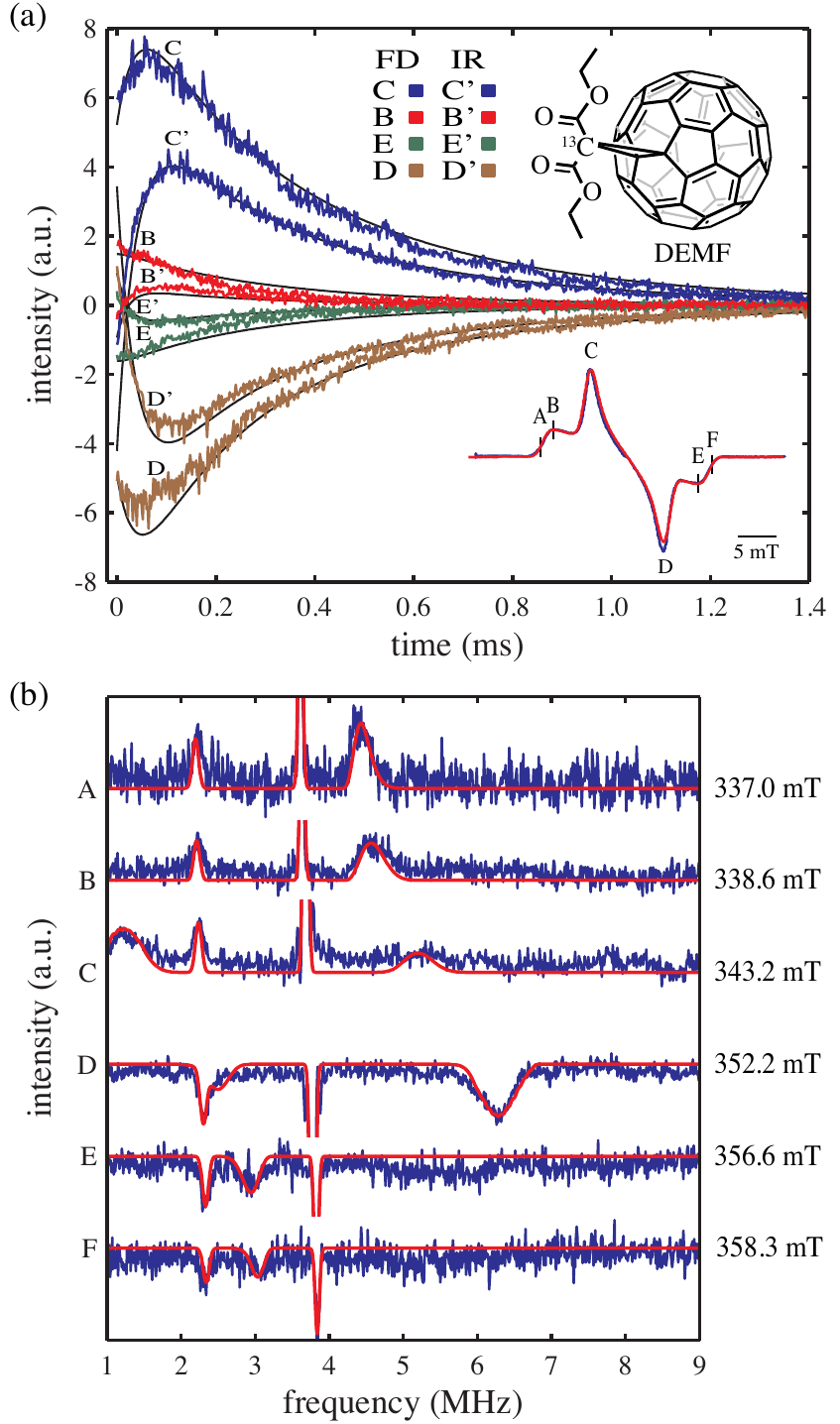}
  \caption{(Color online) (a) Decay traces and simulations of the flash delay (FD) and inversion recovery (IR) experiments at \unit{20}{\kelvin}. Upper inset: DEMF molecule. Lower inset: electron spin echo field sweep of DEMF at \unit{20}{\kelvin}. (b) Mims ENDOR simulations and experimental data acquired at \unit{20}{\kelvin} at fields A-F from (a). The largest peak, around \unit{3.9}{\mega\hertz}, was found to correlate to the $^{13}$C-$T_0$ interaction, while the weaker peak, at \unit{2.2}{\mega\hertz}, related to $^2$H-$T_0$ interaction from the solvent (toluene-d8). Also observable in the ENDOR spectrum is the field dependent component of the hyperfine from the $^{13}$C-$T_{\pm}$ triplet levels.}
  \label{fig:epr}
\end{figure}

There are many molecules possessing optically-excited triplet states which couple to nearby nuclear spins. These include pentacene, pyrazine, and $^{13}$C$_{60}$~\cite{Yago.Link.ea2007Pulsedelectronnuclear,Sloop.Yu.ea1981Electronspinechoes,Zimmermann.Schwoerer.ea1975Endoroftriplet,Donckers.Schwencke.ea1992electronnucleardoubleresonance,Berg.Heuvel.ea1998PulsedENDORStudies}. We here determine the suitability of using a functionalized C$_{60}$ molecule for this scheme. Fullerenes possess a reasonable absorption coefficient at  \unit{532}{\nano\meter} and a highly efficient intersystem crossing (ISC) to a hyperpolarized triplet state~\cite{Arbogast.Darmanyan.ea1991Photophysicalpropertiesof,Wasielewski.ONeil.ea1991Tripletstatesof}. Moreover, the number of spin active nuclei in these molecules is small, suppressing unwanted decoherence {\cite{Morton.Tyryshkin.ea2007Environmentaleffectselectron}. The HF couplings between $^{13}$C atoms on the C$_{60}$ cage and the photoexcited triplet are anisotropic with the largest HF term being \unit{167.6}{\mega\hertz}~\cite{Berg.Heuvel.ea1998PulsedENDORStudies,Dauw.Berg.ea2000tripletwavefunction}. However, since the cage distortion in the excited state introduces a random element to the hyperfine coupling, we shall consider a functionalized fullerene whose molecular axis ensures a well-defined interaction. 
 
Ultimately, an entangling operation would require two $^{13}$C spins, such as those of the trans-1 bis-adduct [FIG.~\ref{fig:entangling-power} (a,b)]. Individual addressibility could be achieved with different chemical shifts of the two nuclei (which typically do not exceed 200 parts per million \cite{Levitt2008Spindynamics}) by making a small modification to one of the groups. In this case Fig. ~\ref{fig:entangling-power} (d) reveals that with equal coupling constants $A$ and $A'$ an optimal entangling operation is still achievable. Here we take the first step towards realizing this scheme by studying the mono-functionalized analogue to determine the key parameters: hyper-polarization populations, triplet lifetime, and the HF tensor with a $^{13}$C labeled methano-carbon of the diethyl malonate mono-adduct (DEMF) [FIG.~\ref{fig:epr} (inset)]. 

EPR measurements on the photo-excited paramagnetic state of spin labeled DEMF were performed at X-band (\unit{9.47}{\giga\hertz}) on a Bruker Elexsys580e spectrometer, equipped with a helium-flow cryostat. Photo-excitation was performed using a \unit{532}{\nano\meter} pulsed Nd-YAG laser with \unit{10}{\hertz} repetition rate; \unit{7}{\nano\second} pulse length; \unit{5-10}{\milli\joule/pulse}. The samples were prepared as \unit{4.3\cdot10^{-4}}{M} solutions in toluene-d$_8$, were deoxygenated using five cycles of freeze-pump-thaw, flame sealed under vacuum and flash frozen in liquid nitrogen.

Upon excitation of the fullerene cage to the singlet excited state the system relaxes through ISC to the lowest triplet state which decays back to the singlet ground state in \unit{90-100}{\micro\second} at \unit{20-50}{\kelvin}, as measured by transient absorption. The ISC relaxation mechanism preferentially populates the triplet sublevels ($T_{x,y,z}$) according to the molecular symmetry. In the presence of a magnetic field the triplet sublevels mix in an orientation dependent manner. This means in principle polarization could be achieved by orienting the molecules in a crystal. The electron spin echo-detected spectra revealed no change in the line shape as a function of excitation delay time or temperature in the range \unit{5-50}{\kelvin}. This is in agreement with previous studies showing that at low temperatures, spin lattice relaxation effects are negligible 
\cite{Terazima.Hirota.ea1992Time-resolvedEPRinvestigation}. EPR line shape simulations were performed using the software package Easyspin~\cite{Stoll.Schweiger2006EasySpincomprehensivesoftware}. The $g$ tensor and ZFS parameters were consistent with previous work [$g_{xx}=2.0006, g_{yy}=2.00115, g_{zz}=2.00215$, $D=\unit{-296}{\mega\hertz}$ and $E=\unit{-6}{\mega\hertz}$]~\cite{Bortolus.Prato.ea2004Time-resolvedEPRstudy}. 

EPR line shape analysis yields only relative differences in populations $p_i$ of the triplet sublevels (e.g.\ $(p_x-p_y)/(p_y-p_z)$)~\cite{Pasimeni.Hirsch.ea1997UseofTransient}. To determine absolute populations and triplet decay rates for the three sub-levels we investigated the relaxation kinetics, as shown in Fig.~\ref{fig:epr}. In our analysis we neglect any decay between the sub-levels and consider separate decay rates $k_{x,y,z}$ from each sub-level to the singlet ground state. These decay rates and the initial populations $n_{x,y,z}$ are each properties of the triplet eigenstates at zero applied magnetic field ($T_{x,y,z}$), and so can be used to determine the resulting decay rates and populations at any applied magnetic field by writing the in-field eigenstates ($T_{+,0,-}$) in the appropriate basis~\cite{WONG:1973p6866}}. Using this approach, we simultaneously fit the decay traces of flash delay ($h\nu\text{-T-}\pi/2\text{-}\tau\text{-}\pi$-echo) and inversion recovery ($h\nu\text{-t-}\pi\text{-T-}\pi/2\text{-}\tau\text{-}\pi$-echo) experiments at four fields, using one set of parameters:  $p_x:p_y:p_z=0.46:0.54:0.00$ and $\tau_x= \unit{(0.50 \pm 0.02)}{\milli\second}$, $\tau_y= \unit{(0.58\pm 0.02)}{\milli\second}$, $\tau_z=\unit{(0.020\pm 0.003)}{\milli\second}$ (\unit{20}{\kelvin}), where $\tau_i=1/k_i$. The transient populations and triplet decay rate were found to show little temperature dependence below \unit{50}{\kelvin}. Importantly, the observed hyperpolarization would allow our electron mediated gate to work well, and could be used to initialize the nuclear spins.

The coupling of the triplet electron and the $^{13}$C spin labeled methano-carbon was measured by Mims electron-nuclear double resonance (ENDOR)~\cite{Mims1965PulsedENDORExperiments} at six field lines, see Fig.~\ref{fig:epr}. Through simulations the hyperfine tensor was found to be axial, $A_{xx}=A_{yy}=2.6\pm0.1$, $A_{zz}=0.74\pm0.02$~MHz, though a $20\%$ strain in $A$ across the sample was required to fit the ENDOR linewidth at all fields (orientations) \cite{Stoll.Schweiger2006EasySpincomprehensivesoftware}. This HF tensor is much smaller than that of certain $^{13}$C atoms in the fullerene cage~\cite{Berg.Heuvel.ea1998PulsedENDORStudies}. 

\begin{figure}[h!]
  \centering
  \includegraphics{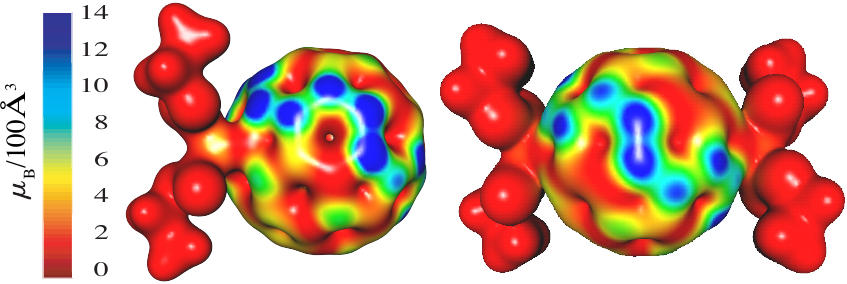}
  \caption{(Color online) (a) Spin density in the lowest triplet excited state of mono- and bis-diethyl malonate fullerenes. The chosen charge isosurface has an isovalue of $0.07 \text{ electron}/\AA^3$ and the overlaid color indicates the magnitude of the spin density. The spin density is mostly gathered around the fullerene cage. Images rendered using Gabedit~\cite{AlloucheGabeditisfree}.}
  \label{fig:FD-DFT}
\end{figure}

In order to understand the weak HF coupling found from ENDOR measurements, and to predict the appropriate coupling of bis-DEMF, we turn to first-principles calculations with density-functional theory (DFT). For this purpose we considered optical excitations to spin triplet states in mono- and bis-diethyl malonate functionalized C$_{60}$ molecules. The calculations were performed within the local density approximation to DFT [see supplementary material for details]. To determine the ground-state structural configuration of the mono-diethyl malonate C$_{60}$ we relaxed three different conformations with symmetries $C_s$, $C_2$, or $C_{2v}$. The molecules with $C_2$ or $C_s$ symmetry were found to be almost degenerate in energy (within \unit{5}{\milli\electronvolt}), and more stable than the molecule with $C_{2v}$ symmetry by \unit{0.32}{\electronvolt}. The excitation to the spin-triplet state was studied in detail for the molecule with $C_2$ symmetry. By carrying out analogous structural relaxations for bis-diethyl malonate we found that the lowest energy configuration carries $C_2$ symmetry. The spin-triplet excited states were calculated by relaxing the electronic structure of each molecule with the total spin constrained to $S_z=1$~\cite{Gunnarsson.Lundqvist1976Exchangeandcorrelation,Kanai.Grossman2007InsightsInterfacialCharge}. Fig.~\ref{fig:FD-DFT} shows that there is significantly more spin polarization on the cage than on the adducts. The calculated HF coupling constants of the $^{13}$C nuclei in mono- and bis-dethyl adduct C$_{60}$ are \unit{4}{\mega\hertz} and \unit{2.2}{\mega\hertz}, respectively. These values are in good agreement with our relatively small measured HF coupling. 

\begin{figure}[h!]
  \centering
  \includegraphics{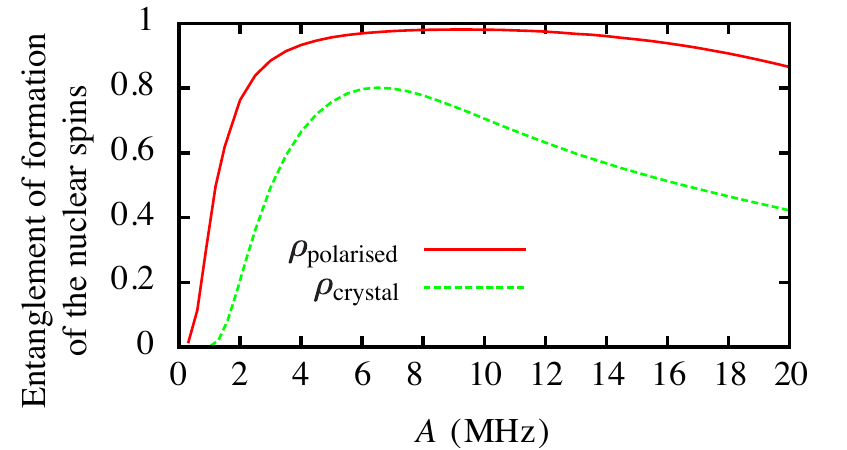}  
 \caption{Entanglement of formation \cite{Wootters1998EntanglementofFormation} of the nuclear spins after applying our scheme and when the excitation has decayed as a function of the HF strength for different initial polarization of the excitation. The nuclear spins for both curves are assumed to be initially in the state $|\downarrow\uparrow\rangle$; parameters as in Fig.~\ref{fig:entangling-power}. For a fully polarized electronic excitation, the simple protocol as described on the first page is applied (solid); lifetimes as below. (Dashed) Here the initial triplet populations based on our experimental findings (see text) are $p_{-}=0.49, p_{0}=0.02, p_{+}=0.49$ with associated lifetimes $\tau_{-}=\unit{0.57}{\milli\second}, \tau_0=\unit{0.02}{\milli\second}, \tau_+=\unit{0.57}{\milli\second}$. In this case, the protocol is applied twice in succession, first for the population in $|T_+ \rangle$ and then for the population in $|T_-\rangle$.}
\label{fig:EntanglementFormation}
\end{figure}

Our experiments on the DEMF molecule provide example triplet relaxation times and hyperpolarization of triplet sub-levels from which we can deduce the strength of HF coupling required to create significant nuclear entanglement [Fig.~\ref{fig:EntanglementFormation}]. When the laser excitation only produces partial electronic polarization, we need to adjust the previously described protocol for creating and preserving entanglement. We consider the  special situation of having all molecules oriented along the $z$-axis with parameters as extracted from the experiments on DEMF, for which the population would be almost equally distributed between $\ket{T_+}$ and $\ket{T_-}$, and the optical decay from $\ket{T_0}$ is much faster than from $\ket{T_{\pm}}$. In this case, we can apply our protocol twice in succession since the entangling power is periodic in time: first for the population initially in $\ket{T_+}$ and, after waiting for  $\ket{T_0}$ to have emptied out, a second time for the population initially in $\ket{T_-}$. Given a two nuclear spin-system exhibiting a hyperfine coupling similar to that observed in our test molecule (\unit{3}{\mega\hertz}), one could perform an operation resulting in a state with entanglement of formation $\approx 0.5$. The hyperfine coupling for the bis-adduct fullerene could be enhanced by using an optimized functional group. In fact, the hyperfine coupling can be regarded as a tunable parameter, and using the results of Fig.~\ref{fig:EntanglementFormation} an optimal value can be chosen to maximize the entangling power. In Fig.~\ref{fig:EntanglementFormation} the entanglement goes down for large coupling constants, because in this case the right-hand side of Eq. \eqref{eq:2} becomes small enough to produce partially distinguishable photons in the decay process.

Our analytic, numerical and experimental results have demonstrated the feasibility of manipulating nuclear spin qubits in molecular structures by harnessing optically excited electron spin states. We identify the required molecular characteristics; it is feasible to synthesise suitable molecules with established techniques. 

\begin{acknowledgments}
\textit{Acknowledgements -} We thank A. Kolli, J. Wabnig and Y. Kanai for stimulating discussions. This work was supported by the Marie Curie Early Stage Training network QIPEST (MESTCT-2005-020505), EPSRC through QIP IRC (GR/S82176/01 and GR/S15808/01), the National Research Foundation and Ministry of Education, Singapore, the DAAD, and the Royal Society.
\end{acknowledgments}


\end{document}